\documentclass[12pt]{iopart}  % preprint
\usepackage{iopams}
\usepackage{setstack}

\def\Jvec{{\bf J}}

\def\Reals{{\mathbb R}}
\def\Complexes{{\mathbb C}}

\begin{document}
\title{Asymptotics of the Wigner $9j$-Symbol}
\author{Hal M. Haggard and Robert G. Littlejohn}
\address{Department of Physics, University of
California, Berkeley, California 94720 USA}

\ead{robert@wigner.berkeley.edu}

\begin{abstract}
We present the asymptotic formula for the Wigner $9j$-symbol, valid
when all quantum numbers are large, in the classically allowed region.
As in the Ponzano-Regge formula for the $6j$-symbol, the action is
expressed in terms of lengths of edges and dihedral angles of a
geometrical figure, but the angles require care in definition.  Rules
are presented for converting spin networks into the associated
geometrical figures.  The amplitude is expressed as the determinant of
a $2\times 2$ matrix of Poisson brackets.  The $9j$-symbol possesses
caustics associated with the fold and elliptic and hyperbolic umbilic
catastrophes.  The asymptotic formula obeys the exact symmetries of
the $9j$-symbol.
\end{abstract}


%\maketitle %use this if you want to have title on separate page

\section{Introduction}

The asymptotic behavior of spin networks has played a significant role
in simplicial approaches to quantum gravity.  Indeed, the field began
with the observation that the Ponzano-Regge action (1968) for the
semiclassical $6j$-symbol is identical to the Einstein-Hilbert action
of a tetrahedron in 3-dimensional gravity in the Regge formulation
(Regge, 1961; see also Williams and Tuckey 1992 and Regge and Williams
2000).  More recently, semiclassical expansions have been used to
study the low-energy or classical limit of quantum gravity as well as
to derive quantum corrections to the classical theory.  Asymptotic
studies in this area have included treatments of the $10j$-symbol
(Barrett and Williams 1999, Baez \etal 2002, Barrett
and Steele 2003, Freidel and Louapre 2003), amplitudes in the
Freidel-Krasnov model (Conrady and Freidel 2008), LQG fusion
coefficients (Alesci \etal 2008), and the EPRL amplitude (Barrett
\etal 2009).  In addition, the venerable $6j$-symbol and Ponzano Regge
(1968) formula continue to receive attention (Roberts 1999, Barrett
and Steele 2003, Freidel and Louapre 2003, Gurau 2008, Charles 2008,
Littlejohn and Yu 2009, Depuis and Livine 2009, Ragni \etal 2010), not
to mention the $q$-deformed $6j$-symbol (Nomura 1989; Taylor and
Woodward 2004, 2005).

In this article we present the generalization of the Ponzano-Regge
formula to the Wigner $9j$-symbol, as well as some material relevant
for the asymptotics of arbitrary spin networks.  The Ponzano-Regge
formula (Ponzano and Regge 1968) gives the asymptotic expression for
the Wigner $6j$-symbol when all quantum numbers are large.  The
$9j$-symbol is the next most complicated spin network after the
$6j$-symbol, with features that are found in all higher spin networks.
In this article we present only the asymptotic formula itself for the
$9j$-symbol and some salient facts surrounding it.  We defer a
derivation and deeper discussion of the formula to a subsequent
publication.

Our derivation has quite a few steps, and some of them at this point
are supported by numerical evidence only.  Thus, we do not now have a
rigorous derivation of our result.  We believe it is correct, however,
on the basis of direct numerical comparisons with the exact
$9j$-symbol, the fact that our formula obeys all the symmetries of the
exact $9j$-symbol, and the plausibility and numerical support for the
conjectures involved in the parts of the derivation currently lacking
proofs.  The proofs do not seem difficult, and we hope to fill in the
gaps in our future work.

Although most of the papers cited above have dealt with the
asymptotics of specific spin networks, usually there are special
values of the angular momenta that are used.  For example, the
$10j$-symbol involves balanced representations of $SO(4)$, which means
that some pairs of $j$'s are equal, while the $9j$-symbols that appear
in LQG fusion coefficients have two columns in which one quantum
number is the sum of the other two.  In addition, $j$'s are sometimes
set equal because this is regarded as the most interesting regime from
a physical standpoint.  

As a result, the spin networks that have been studied tend to fall on
caustics where the asymptotic behavior is not generic.  At such
points, the value of the spin network (the wave function) is not
oscillatory in a simple sense, instead it has the form of a
diffraction catastrophe (Berry 1976).  In addition, the wave function
scales as a higher (less negative) power of the scaling parameter
(effectively, $1/\hbar$).  This type of behavior has been noted in
several places in the quantum gravity literature, although as far as
we can tell no one has noted that it is related to standard caustic
and catastrophe types.  In this article we give a rather complete
picture of the $9j$-symbol for all possible parameters in the
classically allowed region, including all phases and Maslov indices.
We also indicate the subsets upon which the behavior is nongeneric and
described by various types of caustics.  We believe that this is the
first time that such information has been available for any spin
network more complicated than the $6j$-symbol.

Another reason for interest in the $9j$-symbol is that it is the
nontrivial part of the Clebsch-Gordan coefficient for $SO(4)$.

Basic references on the Wigner $9j$-symbol include Edmonds (1960),
Biedenharn and Louck (1981ab) and Varshalovich \etal (1981).  Recent
work on the $9j$-symbol has included new asymptotic forms when some
quantum numbers are large and others small (Anderson \etal 2008,
2009).  We also note the use of $SU(2)$ spin networks in quantum
computing (Marzuoli and Rasetti 2005).

In Sec.~\ref{asymptoticformula} we present the asymptotic formula for
the $9j$-symbol and draw comparisons with the Ponzano-Regge formula to
introduce its geometrical content.  A detailed explanation of the
notation follows in later sections.   In Sec.~\ref{tog} we present
general rules for converting spin networks into surfaces composed of
oriented edges and oriented triangles, and illustrate them for the
$9j$-symbol.  In Sec.~\ref{findingvectors} we explain how the
geometrical objects (pieces of oriented surfaces) corresponding to the
$9j$-symbol can be constructed in 3-dimensional space.  In
Sec.~\ref{visualizing} we explain the configuration space of the
$9j$-symbol and the classically allowed subset thereof.  In
Sec.~\ref{theamplitude} we define the amplitude of the asymptotic
formula and discuss the manifolds (the caustics) upon which it
diverges as well as the diffraction catastrophes that replace the
simple asymptotic form in the neighborhood of the caustics.  In
Sec.~\ref{thephase} we explain the phase of the semiclassical
approximation, a generalization of the Ponzano-Regge action that
requires careful definitions of dihedral angles.  In
Sec.~\ref{symmetries} we show that the asymptotic formula correctly
obeys the symmetries of the $9j$-symbol.  Finally, in
Sec.~\ref{conclusions} we present some comments and conclusions.

\section{The Asymptotic Formula}
\label{asymptoticformula}

The asymptotic expression for the $9j$-symbol is
	\begin{equation}
	\left\{\begin{array}{ccc}
	j_1 & j_2 & j_3 \\
	j_4 & j_5 & j_6 \\
	j_7 & j_8 & j_9
	\end{array}\right\} = A_1 \cos S_1 + A_2 \sin S_2,
	\label{theresult}
	\end{equation}
where $A_{1,2}$ are positive amplitudes, $S_{1,2}$ are phases,
and each term is roughly similar to the single term in the
Ponzano-Regge formula for the $6j$-symbol.  The right hand side is the
leading term in an asymptotic expansion in powers of $1/k$ of the
$9j$-symbol when all nine $j$'s are scaled by a positive factor $k$
that is allowed to go to infinity ($k$ plays the role of $1/\hbar$ in
the asymptotic expansion).  The $k$'s are suppressed in
(\ref{theresult}), but the expression on the right scales as $1/k^3$.
Equation~(\ref{theresult}) applies only in the classically allowed
region.  We do not present the analog of (\ref{theresult}) in the
classically forbidden region.

Equation~(\ref{theresult}) breaks down near caustics, where the
$9j$-symbol scales with a higher (less negative) power of $k$ than
$1/k^3$.  In the neighborhood of caustics, the $9j$-symbol is
approximated by diffraction catastrophes, including the fold and
hyperbolic and elliptic umbilic.  These are discussed more fully in
Sec.~\ref{theamplitude}.

To explain the meaning of (\ref{theresult}) some analogies with the
Ponzano-Regge formula for the $6j$-symbol are useful.  In the
classically allowed region, the Ponzano-Regge formula associates a
given $6j$-symbol with a real tetrahedron whose six edge lengths are
$J_i = j_i+1/2$, where the six $j$'s are those appearing in the
$6j$-symbol.  More precisely, there are two tetrahedra, related by
spatial inversion, that is, time-reversal.  Except for flat
configurations, the two tetrahedra are not related by proper rotations
in $SO(3)$.  We recall that time-reversal, not parity, inverts the
direction of angular momentum vectors.  The two tetrahedra correspond
to the two stationary phase points of the $6j$-symbol, which make
contributions to the asymptotic expression that are complex conjugates
of each other.  The result is the real cosine term in the
Ponzano-Regge formula.  One can say that semiclassically the
$6j$-symbol is a superposition of two amplitudes, corresponding to a
tetrahedral geometry and its time-reversed image, that produce
oscillations in the result.

We shall use lower case $j$'s for quantum numbers, and capital $J$'s
for the lengths of the corresponding classical vectors.  These are
always related by $J_i=j_i+1/2$.  The $1/2$ is a Maslov index (Maslov
and Fedoriuk 1981, Mischenko \etal 1990, de Gosson 1997), and the
manner in which it arises in this context is explained in Aquilanti
\etal (2007).

In the case of the $9j$-symbol in the classically allowed region,
there are four geometrical figures associated with a given set of nine
$j$'s, consisting of two pairs related by time-reversal.  The four
geometrical figures correspond to the four real stationary phase
points of the $9j$-symbol.  Each pair of figures is associated with an
``admissible'' root (defined momentarily) of a certain quartic
equation.  There are two admissible roots in the classically allowed
region, labeled 1 and 2, corresponding to the two terms in
(\ref{theresult}).  Each trigonometric term in (\ref{theresult})
consists of an exponential and its complex conjugate, corrresponding
to a geometrical figure and its time-reversed image.  One can say that
semiclassically the $9j$-symbol is a superposition of four amplitudes
corresponding to four geometries, consisting of two pairs of a
geometry and its time-reversed image.  We now explain these geometries
and how they are specified by the nine $j$'s that appear in the
symbol.

\section{Triangles, Orientations and Geometries}
\label{tog}

The $9j$-symbol specifies the lengths $J_i=j_i+1/2$ of nine classical
angular momentum vectors $\Jvec_i$ but not their directions.
Therefore we inquire as to how the directions may be determined, and
geometrical figures constructed out of the resulting vectors.

Actually, it is convenient to double this set and speak of 18
classical vectors $\Jvec_i$, $\Jvec'_i$, $i=1,\ldots,9$.  A doubling
of this kind was introduced by Roberts (1999), who gave a highly
symmetrical way of writing the $6j$-symbol as a scalar product in a
certain Hilbert space.  Although Roberts only worked with the
$6j$-symbol, his method is easily generalized to an arbitary spin
network.  Ponzano and Regge (1968) also gave hints that doubling of
angular momentum vectors are important in the asymptotic analysis of
spin networks.

We now describe rules that take an arbitary spin network (with at most
trivalent vertices) and transcribe it into relations among a doubled
set of classical angular momentum vectors, defining a set of oriented
triangles and oriented edges of a geometrical figure.  We exemplify
these rules only in the case of the $9j$-symbol, but they are easily
applied to any spin network.  The reader may find it illuminating to
apply our rules to the $6j$-symbol, starting with the usual spin
network (the Mercedes graph).  Figure~\ref{network} illustrates the
spin network of the $9j$-symbol.  See also Fig.~18.1 of Yutsis \etal
(1962).

Each edge of the spin network, labeled by $j_i$, is associated with
two classical angular momentum vectors $\Jvec_i$ and $\Jvec'_i$ that
are required to satisfy
	\begin{equation}
	|\Jvec_i|=|\Jvec'_i| = J_i=j_i+1/2
	\label{lengtheqn}
	\end{equation}
and	  
	\begin{equation}
	\Jvec_i+\Jvec'_i=0.
	\label{JJprimeeqn}
	\end{equation}
Vectors $\Jvec_i$ and $\Jvec'_i$ have the same length and point in
opposite directions.

Each vertex of the spin network, where three edges meet, corresponds
to three vectors that add to zero.  The three vectors are associated
with the three edges.  If the arrow on an edge ending at the vertex is
pointing away from the vertex, then the angular momentum vector is
unprimed; if it is pointing toward the vertex, then the vector is
primed.  This rule applied to Fig.~\ref{network} gives
	\begin{equation}
	\eqalign{
	\Jvec_1 + \Jvec_2 + \Jvec_3 &=0, \qquad
	\Jvec'_1 + \Jvec'_4 + \Jvec'_7=0, \\
	\Jvec_4 + \Jvec_5 + \Jvec_6 &=0, \qquad
	\Jvec'_2 + \Jvec'_5 + \Jvec'_8 =0, \\
	\Jvec_7 + \Jvec_8 + \Jvec_9 &=0, \qquad
	\Jvec'_3 + \Jvec'_6 + \Jvec'_9=0.}
	\label{triangleeqns}
	\end{equation}
These are a set of classical triangle relations, one for each vertex
of the spin network.  In the case of the $9j$-symbol, they are
obviously related to the rows and columns of the symbol.

Although the vector addition in (\ref{triangleeqns}) is commutative,
we agree to write the vectors in each equation in counterclockwise
order (around the vertex of the spin network) for a vertex with $+$
orientation, and in clockwise order for a vertex with $-$ orientation,
modulo cyclic permutations.  Thus the ordering of the vectors is the
same as the ordering of the columns of the $3j$-symbol implied by the
vertex of the network.

This ordering is used to define a set of oriented triangles.  We take
the three vectors of any one of the equations (\ref{triangleeqns}) and
place the base of one vector at the tip of the preceding one, to
create the three edges of a triangle.  In this process we parallel
translate the vectors (in $\Reals^3$) but do not rotate them.  The
triangle is given an orientation (a definition of a normal) by taking
the cross product of any two successive vectors defining the
edges. For example, the normal to the 123-triangle is
$\Jvec_1\times\Jvec_2$, and that of the $1'4'7'$-triangle is
$\Jvec'_1\times\Jvec'_4$, which, in view of (\ref{JJprimeeqn}), is the
same as $\Jvec_1\times\Jvec_4$.

Next, we take the triangles and displace them so that the edge
$\Jvec_i$ of one triangle is adjacent to the edge $\Jvec'_i$ of
another triangle.  In this process, the triangles are displaced but
not rotated.  If we do this with the six triangles defined by
(\ref{triangleeqns}) in the case of the $9j$-symbol, we find that six
pairs of edges can be made adjacent, as illustrated by the central six
triangles of Fig.~\ref{triangles}.  In this ``central region'' six
pairs of vectors $\Jvec_i$ and $\Jvec'_i$ are adjacent for
$i=1,2,5,6,7,9$.  There is some arbitrariness in choosing which six
pairs of edges will be made adjacent.  If we wish that the remaining
edges $i=3,4,8$ also be paired, we can duplicate three of the
triangles and attach them to the periphery of the central region, as
illustrated in Fig.~\ref{triangles}.  This amounts to a kind of
``analytic continuation'' of the central region.

Figure~\ref{triangles} is highly schematic.  In general, the triangles
are not equilateral, the surface that is formed by attaching them
together is not planar, and the triangles may fold under one
another.  

The central region in Fig.~\ref{triangles} is a piece of an oriented
surface, that is, all the normal vectors (by our convention) are
pointing on the same side.  In the case of the $6j$-symbol, our rules
produce a closed surface (the usual tetrahedron), with normals all
pointing either outward or inward (time-reversal converts one into the
other).  In the case of the $9j$-symbol, the surface is not closed.
There is some suggestion that this surface represents a triangulation
of $\Reals P^2$ but for this article we shall view it as living in
$\Reals^3$.

Finally, we orient each edge by choosing the direction of the vector
$\Jvec_i$ (not $\Jvec'_i$).  

We will be interested in finding solutions $\{\Jvec_i, \Jvec'_i,
i=1,\ldots,9\}$ of (\ref{lengtheqn}), (\ref{JJprimeeqn}) and
(\ref{triangleeqns}), modulo overall proper rotations (in $SO(3)$).
That is, although we do not rotate vectors or faces when forming our
surface with oriented faces and egdes, we are allowed to rotate the
whole surface once completed.

We notice that if $\{\Jvec_i, \Jvec'_i, i=1,\ldots,9\}$ is a solution
of these equations, then the time-reversed set $\{-\Jvec_i, -\Jvec'_i,
i=1,\ldots,9\}$ is also a solution.  If we apply our rules for
converting vectors into a surface, we will find in general that the
time-reversed set produces a different surface (not equivalent under
$SO(3)$).  We apply time-reversal only to the vectors, not the rules;
for example, the ordering of the time-reversed vectors is the same as
the original vectors.  The central six triangles of the time-reversed
surface are illustrated in Fig.~\ref{trev}.

To visualize the surfaces in Figs.~\ref{triangles} and \ref{trev}, we
may imagine that the central region of Fig.~\ref{triangles} bulges out
of the paper, like the northern hemisphere of a sphere (whether it
does or not depends on the parameters, but this is one possibility).
Then the time-reversed surface in Fig.~\ref{trev} bulges into the
paper, since spatial inversion is equivalent, modulo $SO(3)$, to
reflection in a plane.  Then the central region of
Fig.~\ref{triangles} can be glued to the time-reversed surface in
Fig.~\ref{trev}, bringing edge $\Jvec'_3$ adjacent to edge $-\Jvec_3$,
etc, and producing a surface homeomorphic to $S^2$.  This is the
hexagonal bipyramid constructed by Ponzano and Regge (1968).  The
conventional normals are pointing outward in the northern hemisphere,
and inward on the southern.  As noted by Ponzano and Regge, this
bipyramid is bisected by three planes passing through a common line,
namely the ``axis'' of the sphere, which cut the bipyramid into three
pairs of congruent tetrahedra.  These correspond to the three
$6j$-symbols in the representation of the $9j$-symbol as a sum over
products of $6j$-symbols (see Edmonds (1960) Eq.~6.4.3), in which the
variable of summation is the common edge of the tetrahedra (the axis
of the sphere).

\section{Finding the Vectors}
\label{findingvectors}

To find a solution of (\ref{lengtheqn}), (\ref{JJprimeeqn}) and
(\ref{triangleeqns}) we notice that all 18 vectors are determined if
only four of them, $\{\Jvec_1,\Jvec_2,\Jvec_4,\Jvec_5\}$ are given.
We let $G$ be the $4\times 4$ Gram matrix constructed out of these
vectors, that is, the $4\times 4$, real symmetric matrix of dot
products of these vectors among themselves.  Of the ten independent
dot products, eight can be determined from the given lengths $J_i$,
$i=1,\ldots,9$.  That is, the diagonal elements are $J_i^2$,
$i=1,2,4,5$, while
	\begin{equation}
	\eqalign{
	\Jvec_1\cdot\Jvec_2 &= (J_3^2-J_1^2-J_2^2)/2,\qquad
	\Jvec_1\cdot\Jvec_4 = (J_7^2-J_1^2-J_4^2)/2,\\
	\Jvec_2\cdot\Jvec_5 &= (J_8^2-J_2^2-J_5^2)/2,\qquad
	\Jvec_4\cdot\Jvec_5 = (J_6^2-J_4^2-J_5^2)/2.}
	\label{dotprods}
	\end{equation}
The two dot products that cannot be determined from the given lengths
are $u=\Jvec_1\cdot\Jvec_5$ and $v=\Jvec_2\cdot\Jvec_4$, which we
regard as unknowns.  These satisfy a linear equation obtained by
squaring $\Jvec_9 = -\Jvec_3-\Jvec_6$, 
	\begin{equation}
	J_9^2=J_3^2+J_6^2 + 2(u+v +\Jvec_1\cdot\Jvec_4
	+\Jvec_2\cdot\Jvec_5).
	\label{xyeqn}
	\end{equation}
Another equation connecting $u$ and $v$ is $\det G=0$, which holds
since the four vectors lie in $\Reals^3$ and the 4-simplex defined by
them is flat.  This is a quartic equation in $u$ and $v$, which by
using (\ref{xyeqn}) to eliminate $v$ can be converted into a quartic
equation in $u$ alone.  We write this quartic as $Q(u)=0$.  We find the
roots $u$ of this quartic, solve for $v$ by using (\ref{xyeqn}),
whereupon all components of the Gram matrix become known (there is one
Gram matrix for each root).

Ponzano and Regge (1968) discussed this procedure in somewhat
different language, and apparently believed that all four roots would
contribute to the asymptotics of the $9j$-symbol.  In fact, they do,
if one wishes to work in the classically forbidden region and/or take
into account tunnelling and exponentially small corrections in the
neighborhood of internal caustic points (more about these below).  But
in the classically allowed region the asymptotics of the $9j$-symbol
are dominated by the contributions from ``admissible'' roots, namely,
those roots that produce Gram matrices that can be realized as dot
products of real vectors $\Jvec_i$.  Only these correspond to real
geometrical figures of the type we have described.

If a root $u$ of $Q(u)=0$ is complex, then it produces a complex Gram
matrix that cannot be realized with real vectors, and so $u$ is
inadmissible.  But a real Gram matrix can be realized as the dot
products of real vectors if and only if it is positive semidefinite, so
even if $u$ is real it will still be inadmissible if $G$ has negative
eigenvalues.   

We define the classically allowed region of the $9j$-symbol as the
region in which $Q(u)$ has at least one admissible root.  In fact, in
the classically allowed region $Q(u)$ has four real roots of which two
are generically admissible.  We order the four real roots of $Q(u)$ in
the classically allowed region in ascending order and label them by
$k=0,1,2,3$.  It turns out that the two admissible roots are the
middle two, $k=1,2$, corresponding to the two terms of
(\ref{theresult}) with the same subscripts, $k=1,2$.

For a given admissible root, that is, a positive semidefinite Gram
matrix, we wish to find the vectors $\Jvec_i$, $i=1,2,4,5$.  We
arrange the four unknown vectors as the columns of a $3\times4$ matrix
$F$, so that $G=F^TF$.  To find $F$ given $G$, we diagonalize $G$,
$G=VKV^T$, where $V\in O(4)$ and $K$ is diagonal with nonnegative
diagonal entries (the eigenvalues of $G$).  At least one of these
eigenvalues must be 0; we place it last, and write $K=D^TD$ where $D$
is a real, $3\times 4$ diagonal matrix.  Then $F=UDV^T$, where $U$ is
an arbitrary element of $O(3)$.  This generates all possible sets of
vectors whose dot products are realized in $G$; it amounts to using
the singular value decomposition of $F$.  If $U=R\in SO(3)$ then we
generate a set of surfaces related by overall rotations; if $U=-R$ we
generate the time-reversed set.  In this way a single Gram matrix,
corresponding to a single admissible root of the quartic, produces a
geometry and its time-reversed image.  Altogether, the two admissible
roots imply the four geometries in (\ref{theresult}).

This method of finding $F$ is discussed in the context of the
$6j$-symbol by Littlejohn and Yu (2009), where it is also applied in
the classically forbidden region.  There we find complex angular
momentum vectors that satisfy the required algebraic relations.  This
carries over to the $9j$-symbol in the classically forbidden region.
In the literature on the $6j$-symbol it is common to state that a
Euclidean group applies in the classically allowed region and a
Lorentz group in the classically forbidden region; but for the
$9j$-symbol the groups are actually $SO(3,\Reals)$ and
$SO(3,\Complexes)$.

\section{The Classically Allowed Region and Configuration Space}
\label{visualizing}

The classically allowed region is a subset of full dimensionality of
the 9-dimensional parameter space of the $9j$-symbol, itself a convex subset
of $\Reals^9$ defined by the triangle inequalities.  To visualize this
and other subsets of the parameter space it helps to fix seven of the
$j$'s to obtain a 2-dimensional slice.  Figure~\ref{configspace}
illustrates such a slice for the case
	\begin{equation}
	\left\{ \begin{array}{ccc}
	129/2 & 137/2 & j_3 \\
	113/2 & 121/2 & j_6 \\
	64 & 108 & 90
	\end{array},
	\right\}
	\label{slice}
	\end{equation}
in which only $j_3$ and $j_6$ are allowed to vary.  The choice of
$j_3$ and $j_6$ for this purpose is not arbitrary, since these two
$j$'s are quantum numbers for a pair of commuting operators on a
space of 5-valent $SU(2)$ intertwiners.  They are like $x$ and $y$ for
a wave function $\psi(x,y)$.  In this analogy, we think of
$(j_3,j_6)$-space as a ``configuration space'' for the $9j$-symbol and
the $9j$-symbol itself as a ``wave function'' $\psi(j_3,j_6)$.
We will mostly use the variables $J_3=j_3+1/2$, $J_6 = j_6+1/2$ to
describe this space.  When thinking in classical terms, $J_3$ and
$J_6$ are continuous variables (not quantized).

Figure~\ref{configspace} illustrates a convex region of the
$J_3$-$J_6$ plane, bounded by straight lines and defined by the
classical triangle inequalities,
	\begin{equation}
	\eqalign{
	\max(|J_1-J_2|,|J_6-J_9|) &\le J_3 \le 
	\min(J_1+J_2,J_6+J_9)\\
	\max(|J_4-J_5|,|J_3-J_9|) &\le J_6 \le
	\min(J_4+J_5,J_3+J_9).}
	\label{triangleinequals}
	\end{equation}
Properly speaking, configuration space is this convex region, not the
whole plane.  The unshaded area inside the convex region is the
classically allowed region, surrounded by the shaded classically
forbidden region.  The caustic curve separates the classically allowed
from the classically forbidden regions; it has kinks (discontinuities
in slope) at points $B$, and is tangent to the boundary of the convex
region at several points.  Other features of this figure are explained
below.

Given a point $(J_3,J_6)$ of the classically allowed region, the
procedure described in Secs.~\ref{tog} and \ref{findingvectors}
produces a quartic polynomial $Q(u)$ whose two middle roots $k=1,2$
are admissible.  These can be thought of as specifying a two-branched
``root surface'' that sits over the classically allowed region.  The
two middle roots coalesce as we approach the caustic curve, and become
(inadmissible) complex conjugates as we move beyond.  Thus, the two
root surfaces can be thought of as being glued together on the caustic
curve.

Corresponding to each root there are two geometries modulo $SO(3)$,
related by time-reversal, so there is a two-fold ``geometry surface''
sitting above each root surface, or four geometry surfaces sitting
above the classically allowed region.  These four geometry surfaces
are actually branches of the projection of an invariant 2-torus onto
configuration space, and correspond to the four exponential terms in
(\ref{theresult}).  This 2-torus sits in the phase space of the
$9j$-symbol, a 4-dimensional, compact symplectic manifold.

This symplectic manifold is only one of several phase spaces that
describe the classical mechanics of the $9j$-symbol, but all the others
have higher dimensionality so we call this one the ``phase space of
minimum dimensionality.''  It is one of the symplectic manifolds
discovered by Kapovich and Millson (1996).  Its analog in the case of
the $6j$-symbol is a spherical phase space, which has been studied by
Charles (2008) and by Littlejohn and Yu (2009).  The phase space of
minimum dimensionality is related to other phase spaces for the
$9j$-symbol by a combination of symplectic reduction (Marsden and
Ratiu 1999) and the elimination of constraints.  We have found it
useful to employ all these spaces in our work on the $9j$-symbol.

\section{The Amplitude and Caustics}
\label{theamplitude}

The amplitudes of semiclassical approximations are notorious for the
computational difficulties they cause.  For example, several authors
have resorted to computer algebra and/or numerical experimentation to
check the amplitude determinant in the Ponzano-Regge formula.
Actually, this amplitude (due originally to Wigner (1959)) is given by
a single Poisson bracket between intermediate angular momenta
(Aquilanti \etal 2007, and, in more detail, Littlejohn and Yu 2009),
which can be evaluated in a single line of algebra.  More generally,
semiclassical amplitudes are easily found in terms of matrices of
Poisson brackets.

In the case of the $9j$-symbol we define
	\begin{equation}
	V_{ijk} = \Jvec_i \cdot (\Jvec_j \times \Jvec_k),
	\label{Vijkdef}
	\end{equation}
which is six times the signed volume of the tetrahedron specified by
edges $i$, $j$, $k$ (it is the volume of the corresponding
parallelepiped).  Then the amplitudes $A_1$, $A_2$ in
(\ref{theresult}) are given by
	\begin{equation}
	A= \frac{1}{4\pi \sqrt{|\det D|}},
	\label{Adef}
	\end{equation}
where
	\begin{equation}
	D= \left(\begin{array}{cc}
	V_{124} & V_{215} \\
	V_{451} & V_{542}
	\end{array}\right).
	\label{Ddef}
	\end{equation}
The subscripts 1,2 are omitted on $A$ in (\ref{Adef}) because the same
formula applies for both terms in (\ref{theresult}), but $A_1 \ne A_2$
in general because the formula is evaluated on two different
geometries (associated with the two admissible roots).  The quantitity
$\det D$ is even under time-reversal, so the same amplitude applies to
both a geometry and its time-reversed image.

The volumes in matrix $D$ are Poisson brackets of intermediate angular
momenta in a recoupling scheme for the $9j$-symbol, which are most
easily evaluated in the phase space of minimum dimensionality.  We
omit details; suffice it to say for now that the derivation of the
matrix (\ref{Ddef}) in terms of Poisson brackets and thence the
amplitude is extremely easy.

We define the caustic set as the subset of the $9j$-parameter space
where $\det D=0$.  Its intersection with the 2-dimensional slice seen
in Fig.~\ref{configspace} consists of the union of the caustic curve
(the curve separating the classically allowed from the classically
forbidden region) with the two points marked $I$.  In addition, the
caustic set includes the continuation of the caustic curves from
points $B$ into the classically forbidden region.  The points $I$ are
``internal'' caustics, that is, internal to the classically allowed
region.  While the caustic curve has codimension 1, the internal
caustics have codimension 2.

The quantity $\det D$ is nonzero away from the caustics.  It turns out
that the sign of $\det D$ distinguishes the two root surfaces, with
$\det D >0$ on root surface 1 and $\det D <0$ on root surface 2.

The caustics of the $6j$-symbol occur at the flat configurations (flat
tetrahedra), as appreciated by Ponzano and Regge (1968) and Schulten
and Gordon (1975a,b).  The caustics of the $9j$-symbol, however, are
not in general flat, that is, $\det D=0$ does not imply that the
configuration is flat.  The flat configurations of the $9j$-symbol,
however, do lie on the caustic set.  In a given $J_3$-$J_6$ slice,
there are precisely four flat configurations.  In the example of
Fig.~\ref{configspace}, these are marked $B$ and $I$.  The points $B$
are flat configurations lying on the boundary of the classically
allowed region (the caustic curve), while points $I$ are internal flat
configurations.  As we vary the seven $j$'s that are fixed in
Fig.~\ref{configspace}, the number of flat configurations on the
boundary varies from 2 to 4; those not on the boundary are internal.

In the usual manner of semiclassical approximations, (\ref{theresult})
breaks down in a neighborhood of the caustic set (it diverges exactly
at the caustic), and must be replaced by a diffraction function
associated with a catastrophe (Berry 1976).  In the case of the
$6j$-symbol, the only catastrophe that occurs is the fold, yielding an
Airy function as the semiclassical approximation, as noted by Ponzano
and Regge (1968) and Schulten and Gordon (1975).  This is the normal
situation for systems of one degree of freedom.  The $9j$-symbol,
however, possesses two degrees of freedom, and other types of
catastrophes occur.  The fold catastrophe applies at most points along
the caustic curve, where the $9j$-symbol is approximated by an Airy
function; but at flat configurations there is an umbilic catastrophe,
hyperbolic for those ($B$) falling on the boundary (caustic) curve and
elliptic for the internal caustics ($I$).  See Trinkhaus and Dreper
(1977) for illustrations of the associated diffraction functions.  The
umbilic catastrophes are generic in systems of three degrees of
freedom but occur in the $9j$-symbol (with only two) because of
time-reversal symmetry.  However, only sections of the full
three-dimensional umbilic wave forms appear (Berry 1976).  The cusp
catastrophe, which can be expected in generic systems of two degrees
of freedom, does not occur in the classically allowed region of the
$9j$-symbol.

Caustics are associated with the coalescence of branches of the
projection of a Lagrangian manifold in phase space onto configuration
space.  In the case of the $9j$-symbol, the Lagrangian manifold is the
invariant 2-torus mentioned in Sec.~\ref{visualizing}.  Along the
boundary of the classically allowed region, the two admissible roots
coalesce, which means that the four geometries merge into two.  At
most points on the boundary curve, the two remaining geometries are
not equal, but are related by time-reversal.  At such points we have a
fold catastrophe, and the $9j$-symbol is approximated by an Airy
function (modulated by a cosine term).  At points $B$, however, the two
geometries related by time-reversal merge into a single flat
configuration, producing the hyperbolic umbilic catastrophe.

At internal caustic points $I$ the geometry and its time reversed
image for one of the two admissible roots coalesce to produce a flat
configuration.  The two geometries of the other root surface, however,
do not coalesce.  Thus at internal caustics $I$ there are three
geometries.  Only the flat configuration associated with one of the
roots produces the elliptic umbilic catastrophe; thus, only one of the
two terms in (\ref{theresult}) is replaced by the elliptic umbilic
diffraction function, while the other remains as shown in
(\ref{theresult}).  The $9j$ symbol is a linear combination of these
two terms, but the elliptic umbilic diffraction function dominates
when the scaling factor $k$ is large.

The caustics have a certain size, that is, a distance around the caustic
set over which diffraction functions must be used instead of
(\ref{theresult}).  This distance $\Delta j$ scales as $k^{1/3}$ for
all three catastrophe types (fold and elliptic and hyperbolic umbilic)
discussed here.  

In the neighborhood of fold catastrophes the wave function scales as
$k^{-17/6}$, that is, $k^{1/6}$ higher than the $k^{-3}$ of the two
terms in (\ref{theresult}).  In the neighborhood of umbilic
catastrophes the scaling is $k^{-8/3}$, that is, with another factor
of $k^{1/6}$.  For large values of $k$ the $9j$-symbol is largest near
the points $I$, $B$.

Linear combinations with different scaling behaviors have been observed
by Barrett and Steele (2003) and by Freidel and Louapre (2003) in
their studies of the $10j$-symbol.  It seems that the $9j$-symbol is
the simplest spin network in which this phenomenon occurs.

\section{The Phase}
\label{thephase}

The phases $S_1$ and $S_2$ in (\ref{theresult}) each have the form
	\begin{equation}
	S = \sum_{i=1}^9 J_i \theta_i,
	\label{Sdef}
	\end{equation}
where $\theta_i$ is the angle between normals of adjacent faces of the
geometrical figure.  This of course is similar to the Ponzano-Regge
formula, but the $6j$-tetrahedron is convex and all dihedral angles
can be taken in the interval $[0,\pi]$.  The dihedral angles for the
$9j$-symbol, on the other hand, must be allowed to lie in a full
$2\pi$ interval, as explained momentarily.  The subscripts 1,2 are
omitted on $S$ in (\ref{Sdef}) because the same formula applies to
both terms in (\ref{theresult}).  The formula must be evaluated,
however, on two different geometries, so $S_1$ and $S_2$ are not
equal.  In addition, the angles $\theta_i$ lie in different intervals
for the two geometries.

Each edge $i$ of the geometrial figure is adjacent to two faces, for
example, edge 4 in Fig.~\ref{triangles} is adjacent to faces $1'4'7'$
and $456$.  One face adjacent to edge $i$ contains vector $\Jvec_i$,
and the other $\Jvec'_i$.  Let the two normals of these two faces,
according to the conventions given above, be $\hat{\bf n}$ and
$\hat{\bf n}'$.  Then we define $\theta_i$ as the angle such that
	\begin{equation}
	R(\hat{\bjmath},\theta_i)\hat{\bf n} = \hat{\bf n'},
	\label{thetadef}
	\end{equation}
where $\hat{\bjmath}$ is the unit vector along $\Jvec$, specifying the
axis of a rotation $R$ by angle $\theta_i$ using the right-hand rule.
In the Ponzano-Regge formula one can compute the dihedral angle from
its cosine, but for the $9j$ one must also use the sine of the
angle. That is, (\ref{thetadef}) is equivalent to
	\begin{equation}
	\hat{\bf n}' = \cos\theta_i \,\hat{\bf n}
	+ \sin\theta_i \,\hat{\bjmath} \times \hat{\bf n}.
	\label{altthetadef}
	\end{equation}
This determines $\theta_i$ to within an additive integer multiple of
$2\pi$.  We add the further requirement that for the geometries
associated with the first root (the cosine term in (\ref{theresult})),
$-\pi\le \theta_i < +\pi$, while for the second root (the sine term in
(\ref{theresult})), $0\le\theta_i < 2\pi$.  These ranges for the angle
$\theta_i$ are chosen because they give a continuous branch for the
angle over the two root surfaces.  It turns out that $\theta_i$ never
crosses $\pm \pi$ on the surface for root 1, and it never crosses 0 or
$2\pi$ on the surface for root 2.

The rules given in Secs.~\ref{tog} and \ref{findingvectors} for
converting vectors into surfaces with oriented edges and triangles are
an essential part of the definition of the dihedral angles
$\theta_i$.  It is of interest to see how the angles change when a set
of vectors or the associated geometry is subjected to some symmetry.

Under time-reversal, the orientation of all triangles 
reverses, that is, the normal vectors stay the same but the vectors
defining the edges are inverted.  This means that the angles
$\theta_i$ go into $-\theta_i$ on root surface 1,
while they go into $2\pi-\theta_i$ on root surface 2 (both changes
guarantee that the angles remain within their respective ranges).
Thus $S$ goes into $-S$ on root surface 1 and 
	\begin{equation}
	S \to -S + 2\pi\nu+ 9\pi
	\label{S2map}
	\end{equation}
on root surface 2, where $\nu$ is the integer,
	\begin{equation}
	\nu=\sum_{i=1}^9 j_i.
	\label{nudef}
	\end{equation}
These guarantee that $\cos S_1$ and $\sin S_2$ are invariant under
time-reversal.  Since the same applies to the amplitudes $A_1$ and
$A_2$, one can choose either a geometry or its time-reversed image,
for each root, when evaluating (\ref{theresult}).  

This completes the definition and geometrical interpretation of all
the notation used in (\ref{theresult}).

\section{Symmetries of the $9j$-symbol}
\label{symmetries}

The formula (\ref{theresult}) transforms correctly under the
symmetries of the $9j$-symbol (Varshalovich \etal 1981, Sec.~10.4),
which state that the $9j$-symbol suffers a phase change of $(-1)^\nu$
under odd permutations of rows or columns or under transposition.
Consider, for example, the swapping of the first two columns, and let
$P$ be the permutation of indices, so that $P1=2$, $P2=1$, $P3=3$,
etc.  This maps an old set of nine $j$'s into a new set, and old
quartic $Q(u)=0$ into a new one, etc.  We find that the $u$ root of
the old quartic becomes the $v$ root of the new one, which amounts to
saying that the root 1 surface of the old geometry is mapped into the
root 2 surface of the new one, and vice versa.  Also, the orientations
of the three unprimed triangles reverse, but not those of the primed
ones, causing all nine dihedral angles to be incremented or
decremented by $\pi$ (depending on the range).  If we let $\theta_i$ be
the original angles and ${\tilde\theta}_i$ the new ones, then when
$\theta_i$ is on root surface 1 we find ${\tilde\theta}_{Pi} = \theta_i
+\pi$, which means that the new angle is in the right range since it
is on root surface 2.  Similarly, when $\theta_i$ is on root surface 2
then ${\tilde\theta}_{Pi} = \theta_i-\pi$, which is in the right range
since ${\tilde\theta}_{Pi}$ is on root surface 1.  As a result, when the
original geometry is on root surface 1, we have
	\begin{equation}
	\sum_{i=1}^9 J_i{\tilde\theta}_i = \sum_{i=1}^9 J_i\theta_i 
	+\nu\pi + \frac{9\pi}{2},
	\end{equation}
so that $\sin {\tilde S}_2 = (-1)^\nu \cos S_1$, while if the
original geometry is on root surface 2, we have
	\begin{equation}
	\sum_{i=1}^9 J_i{\tilde\theta}_i = \sum_{i=1}^9 J_i\theta_i
	-\nu\pi -\frac{9\pi}{2},
	\end{equation}
so that $\cos{\tilde S}_1 = (-1)^\nu \sin S_2$.  The sine and cosine
terms in (\ref{theresult}) swap under column swap, and the result
acquires an overall phase of $(-1)^\nu$, as required.  The specified
ranges on the dihedral angles on the two root surfaces are necessary
for this to work out.

\section{Comments and Conclusions}
\label{conclusions}

It is easy to derive the expression (\ref{Sdef}) by the method of
Roberts (1999), which involves rotating faces by an angle of $\pi$
about their normals, and edges by an angle of $\pi$ about a normal to
them.  The phase (\ref{Sdef}) (times 2) is then an action integral
along one Lagrangian manifold and back along another (the analogs of
the $A$- and $B$-manifolds of Aquilanti \etal (2007)).  Similar
expressions apply to any spin network of any complexity.  But the
contours chosen for the integration are not unique, in that one can
add any multiples of quantized loops on the two manifolds.  These
modify both the actions and the Maslov indices, and amount to changing
the choice of branch for the angles $\theta_i$, that is, adding an
integer multiple of $2\pi$ to these angles.  This does not leave the
trigonometric functions in (\ref{theresult}) invariant because the
angles are multiplied by the $J_i$, which may be half-integers.  The
result is that the phase of the approximation to the $9j$-symbol
depends on the contours.  A more serious worry is that the contours,
that is, the branches for the $\theta_i$, may change as we move around
in the parameter space of the $9j$-symbol.  This would amount to
crossing a branch cut for the angles $\theta_i$ (and there are
different branch cuts for different angles).  In addition, as we move
around in parameter space we can make any two adjacent faces rotate
relative to one another around their common edge as many times as we
want.  Although the phases in question are ``only'' powers of $-1$,
straightening out this issue was by far the hardest part of this work.
In the end we realized that the ranges $[-\pi,+\pi)$ on root surface 1
and $[0,2\pi)$ on root surface 2 guarantee that there are no branch
cuts and hence no discontinuities.  The ranges specified for the
angles $\theta_i$ give us in effect a global, smooth definition of
contours for carrying out action integrals.

We present several numerical comparisons of (\ref{theresult}) with the
exact $9j$-symbol.  In Fig.~\ref{comparison} the approximation
(\ref{theresult}) (smooth curve) may be compared to the exact
$9j$-symbol (sticks) as a function of $j_3$ for fixed values of the
other $j$'s.   The range chosen lies inside the classically allowed
region, far from a caustic.  Fig.~\ref{comparefold} shows the
comparison in a range that crosses a fold catastrophe, and
Fig.~\ref{compareumbilic} shows the comparison in an interval that
passes near a hyperbolic umbilic catastrophe (the upper point $I$ in
Fig.~\ref{configspace}).  The approximation (\ref{theresult}) is too
large near the point $I$.

Varshalovich \etal (1981) present an asymptotic approximation for the
$9j$-symbol without citation (their Eq.~(10.7.1)).  We believe their
formula must be an asymptotic expression for the $9j$-symbol in a
different sense than we have defined it, or else it is incorrect.

The two terms in (\ref{theresult}) have different trigonometric
functions (sine and cosine) because there is a relative Maslov index
of 2 between the two root surfaces.  The relative Maslov index between
a geometry and its time-reversed image is 0, a somewhat surprising
result because in mechanical systems and in the $6j$-symbol the Maslov
index between a branch or geometry and its time-reversed image is 1.

When an interior caustic occurs on a root surface, the two geometries
that sit above it form a double cover, in the manner of the Riemann
sheet for the square root function.  The internal caustic point $I$ is
a branch point for the cover.  Geometries transform continuously into
their time-reversed images as we go around the point $I$, without
crossing a caustic.

Several studies of the asymptotics of spin networks have started with
an integral representation of the network, to which the stationary
phase approximation is applied.  Roberts (1999) represented the
$6j$-symbol as a scalar product in a certain Hilbert space, which was
put into the coherent state representation, whereupon the integral was
evaluated by the stationary phase approximation.  Coherent states have
played a prominent role in many recent semiclassical studies.  Our
approach has been to work as much as possible in a
representation-independent manner.  For example, the stationary phase
points are seen as intersections of Lagrangian manifolds.  Some of the
basics of this approach were presented in Aquilanti \etal (2007).  We
have not specifically used the coherent state or any other
representation.

Some aspects of this calculation carry through in an obvious way to
higher spin networks, while for others nontrivial generalizations seem
to be required.  But we believe that an understanding of the $9j$
results are necessary for a full understanding of the asymptotics of
higher spin networks.

We will report in more detail on the derivation of (\ref{theresult})
in a later publication.

\ack

The authors would like to thank Enzo Aquilanti, Mauro Carfora,
Annalisa Marzuoli and Carlo Rovelli for encouragement, many useful
pieces of information, many stimulating conversations, and much warm
hospitality during the progress of this work.  We would also like to
thank Cynthia Vinzant for discussion of positive semidefinite
completion, which helped greatly in proving that there are generically
two admissible roots in the classically allowed region.  

This work was supported by a grant from the France-Berkeley Fund.

\section*{References}
\begin{harvard}

\item[] Alesci E, Bianchi E, Magliaro E and Perini C 2008 preprint
gr-qc 0809.3718 

\item[] Anderson R W, Aquilanti V and Ferreira C da S 2008 {\it
J. Chem. Phys.} {\bf 129} 161101

\item[] Anderson R W, Aquilanti V and Marzuoli A 2009 {\it
J. Phys. Chem. A} {\bf 113} 15106

\item[] Aquilanti V, Haggard H M, Littlejohn R G and Yu L 2007 {\it
J Phys A} {\bf 40} 5637

\item[] Baez J C Christensen J D and Egan G 2002 {\it Class. Quantum
Grav.} {\bf 19} 6489

\item[] Barrett J W, Dowdall R J, Fairbairn W J, Gomes H and Hellmann
F 2009 preprint gr-qc 0902.1170

\item[] Barrett J W and Steele C M 2003 {\it Class. Quantum Grav.}
{\bf 20} 1341

\item[] Barrett J W and Williams R M 1999 {\it Adv. Theor. Math. Phys.}
{\bf 3} 209

\item[] Berry M V 1976 {\it Adv. Phys.} {\bf 25} 1

\item[] Biedenharn L C and Louck J D 1981a {\it Angular Momentum in 
Quantum Physics} (Reading, Massachusetts: Addison-Wesley)

\item[] \dash 1981b {\it The Racah-Wigner Algebra in Quantum Theory}
(Reading, Massachusetts: Addison-Wesley)

\item[] Charles L 2008 preprint math 0806.1585

\item[] Conrady F and Freidel L 2008 {\it Phys. Rev. D} {\bf 78} 104023

\item[] de Gosson M 1997 {\it Maslov Classes, Metaplectic
Representation and Lagrangian Quantization} (Berlin: Akademie Verlag)

\item[] Dupuis M and Livine E R 2009 {\it Phys. Rev. D} {\bf 80} 024035

\item[] Edmonds A R 1960 {\it Angular Momentum in Quantum Mechanics}
(Princeton: Princeton University Press)

\item{} Freidel L and Louapre D 2003 {\it Class. Quantum Grav.} {\bf
20} 1267

\item[] Gilmore R 1981 {\it Catastrophe Theory for Scientists and
Engineers} (New York: Dover)

\item[] Gurau R 2008 {\it Annales Henri Poincar\'e} {\bf 9} 1413

\item[] Kapovich M and Millson J J 1996 {\it J. Differential Geometry}
{\bf 44} 479

\item[] Littlejohn R G and Yu L 2009 {\it J. Phys. Chem. A} {\bf 113} 
14904

\item[] Marsden J E and Ratiu T 1999 {\it Introduction to Mechanics and
  Symmetry} (New York: Springer-Verlag)

\item[] Marzuoli A and Rasetti M 2005 {\it Ann. Phys.} {\bf 318} 345

\item[] Maslov V P and Fedoriuk M V 1981 {\it Semi-Classical
Approximation in Quantum Mechanics} (Dordrecht, Holland: D. Reidel)

\item[] Mishchenko A S, Shatalov V E and Sternin B Yu 1990 {\it Lagrangian
manifolds and the Maslov operator} (Berlin: Springer Verlag)

\item[] Nomura M 1989 {\it J. Math. Phys.} {\bf 30} 2397

\item[] Ponzano G and Regge T 1968 in {\it Spectroscopy and Group
Theoretical Methods in Physics} ed F Bloch \etal\ (Amsterdam:
North-Holland) p~1

\item[] Ragni M, Bitencourt A C P, Ferreira C da S, Aquilanti V, 
Anderson R W and Littlejohn R G 2010 {\it Int. J. Quantum Chem.} 
{\bf 110} 731

\item[] Regge T 1961 {\it Nuovo Cimento} {\bf 19} 558

\item[] Regge T and Williams R M 2000 {\it J. Math. Phys.} {\bf 41} 3964

\item[] Roberts J 1999 {\it Geometry and Topology} {\bf 3} 21

\item[] Schulten K and Gordon R G 1975a {\it J. Math. Phys.} {\bf 16} 1961

\item[] \dash 1975b {\it J. Math. Phys.} {\bf 16} 1971

\item[] Taylor Y U and Woodward C T 2004 preprint math 0406.228

\item[] \dash 2005 {\it Sel. Math. New ser.} {\bf 11} 539

\item[] Trinkhaus H and Drepper F 1977 {\it J. Phys. A} {\bf 10} L11

\item[] Varshalovich D A, Moskalev A N and Khersonskii V K 1981 {\it
Quantum Theory of Angular Momentum} (Singapore:  World Scientific)

\item[] Wigner E P 1959 {\it Group Theory} (Academic Press, New York)

\item[] Williams R M and Tuckey P A 1992 {\it Class. Quantum Grav.} 
{\bf 9} 1409

\item[] Yutsis A P, Levinson I B and Vanagas V V 1962 {\it The Theory of
Angular Momentum} (Jerusalem: Israel Program for Scientific
Translations)

\end{harvard}

\Figures

\begin{figure}
\caption{\label{network} The spin network for the $9j$-symbol.}
\end{figure}

\begin{figure}
\caption{\label{triangles} The six triangles defined by
(\ref{triangleeqns}) form the ``central region'' of the figure, with
three triangles duplicated and attached to the edges of the central
region.  The notation $1$, $2'$, etc refers to $\Jvec_1$, $\Jvec'_2$,
etc.}
\end{figure}

\begin{figure}
\caption{\label{trev} The central region of the time-reversed
surface.  Notation $-1$, $-2'$ etc refers to $-\Jvec_1$, $-\Jvec'_2$,
etc.}
\end{figure}

\begin{figure}
\caption{\label{configspace} The convex region of the $J_3$-$J_6$
plane is the configuration space of the $9j$-symbol.  The shaded area
is the classically forbidden region, and the unshaded, the classically
allowed.  Points $I$ are internal caustics, two of the flat
configurations; points $B$ are the other two flat configurations,
lying on the boundary curve.}
\end{figure}

\begin{figure}
\caption{\label{comparison} Comparison of exact $9j$-symbol (vertical sticks) with approximation (\ref{theresult}), away from a caustic.  Values used are those in (\ref{slice}), with $j_6=50$.}
\end{figure}

\begin{figure}
\caption{\label{comparefold} Like Fig.~\ref{comparison}, but an
interval that spans a fold catastrophe (with $j_6=60$).  The approximation
(\ref{theresult}) is discontinued at the caustic, the exact values are
continued into the classically forbidden region.}
\end{figure}

\begin{figure}
\caption{\label{compareumbilic} Like Fig.~\ref{comparefold}, but
passing near an ellipitc umbilic catastrophe (with $j_6=79$).}
\end{figure}

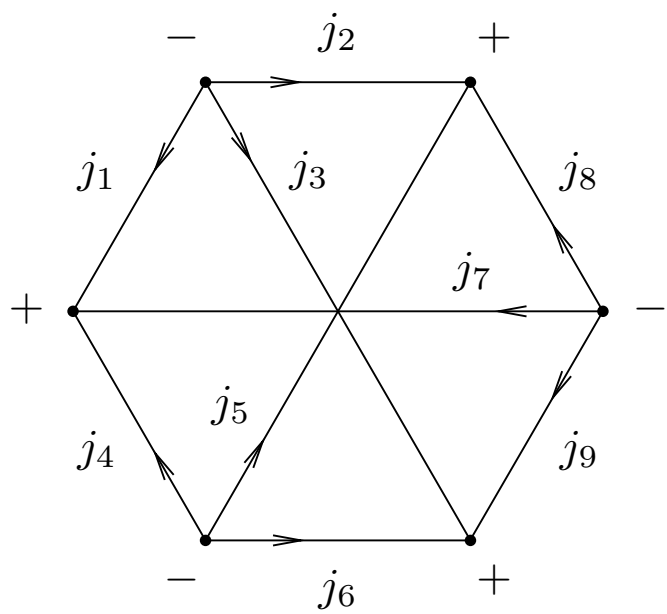
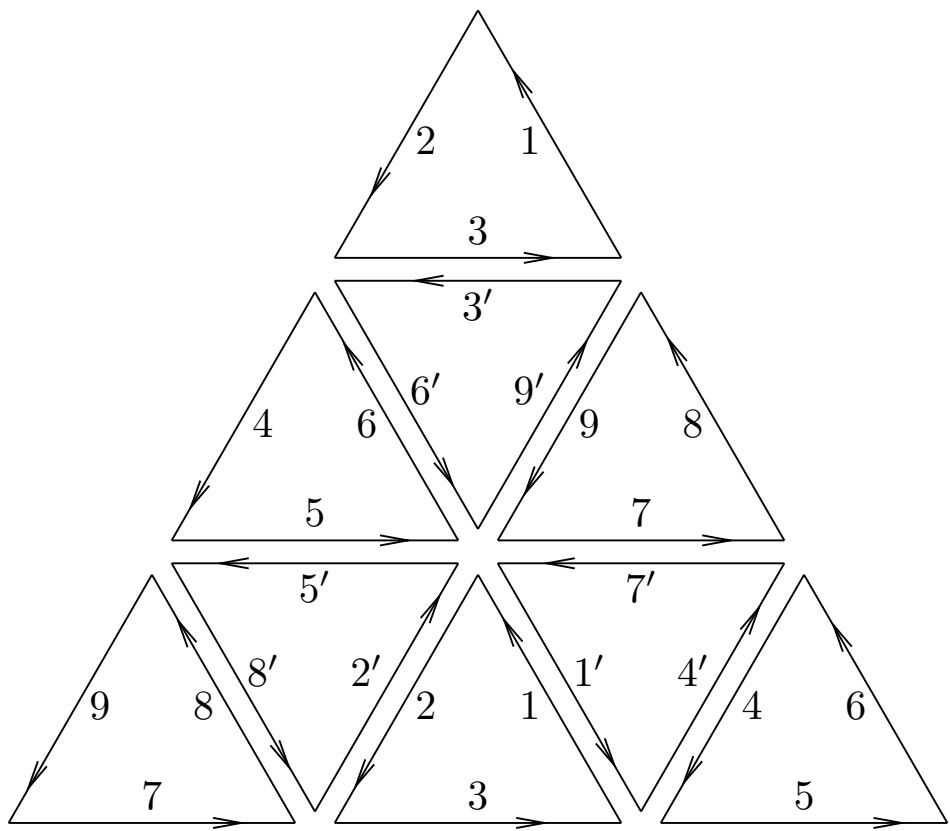
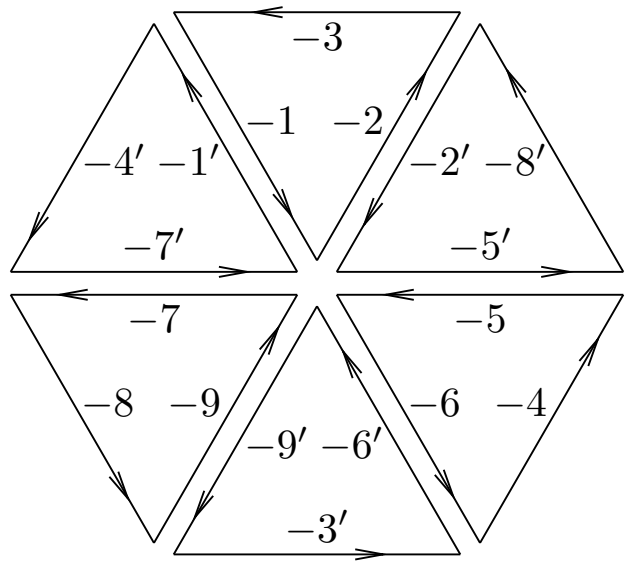
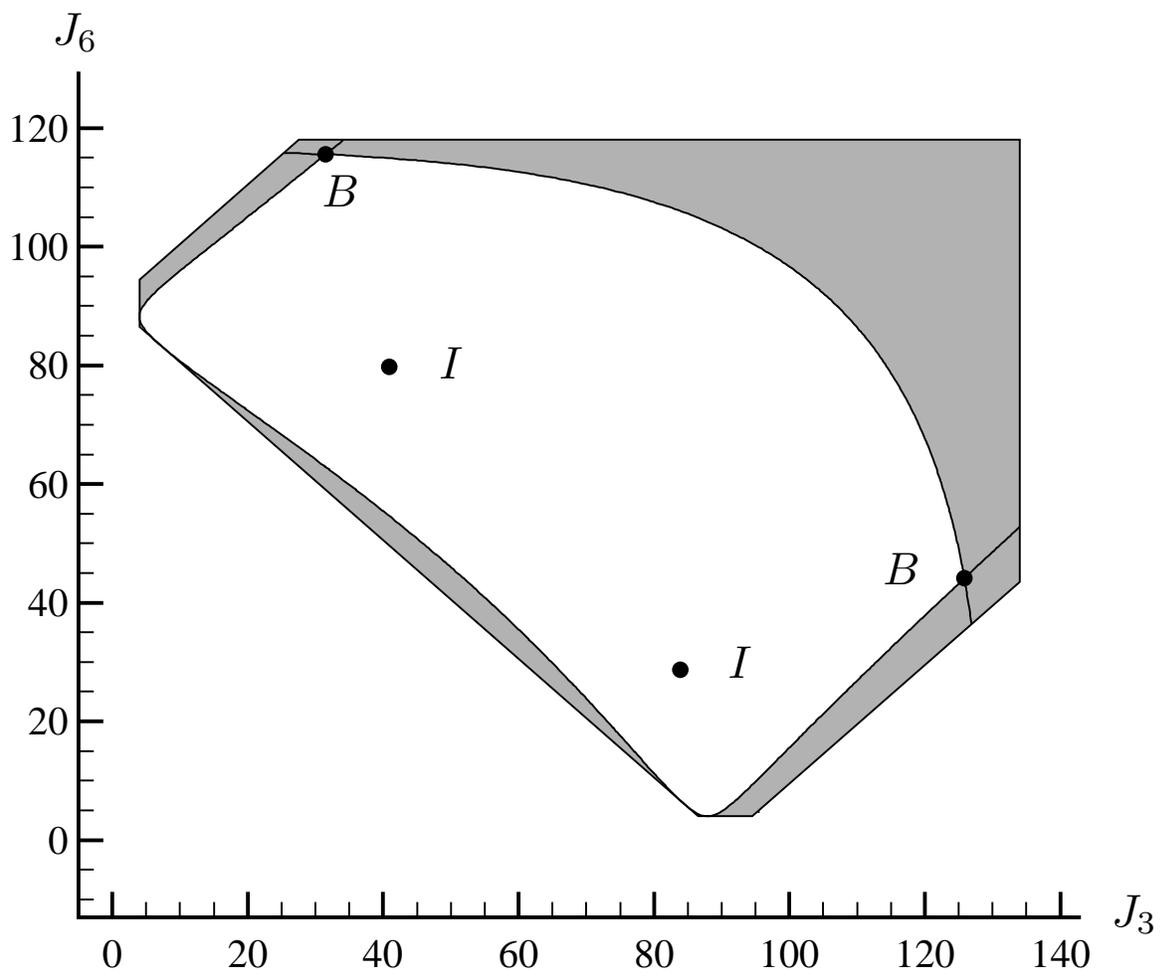
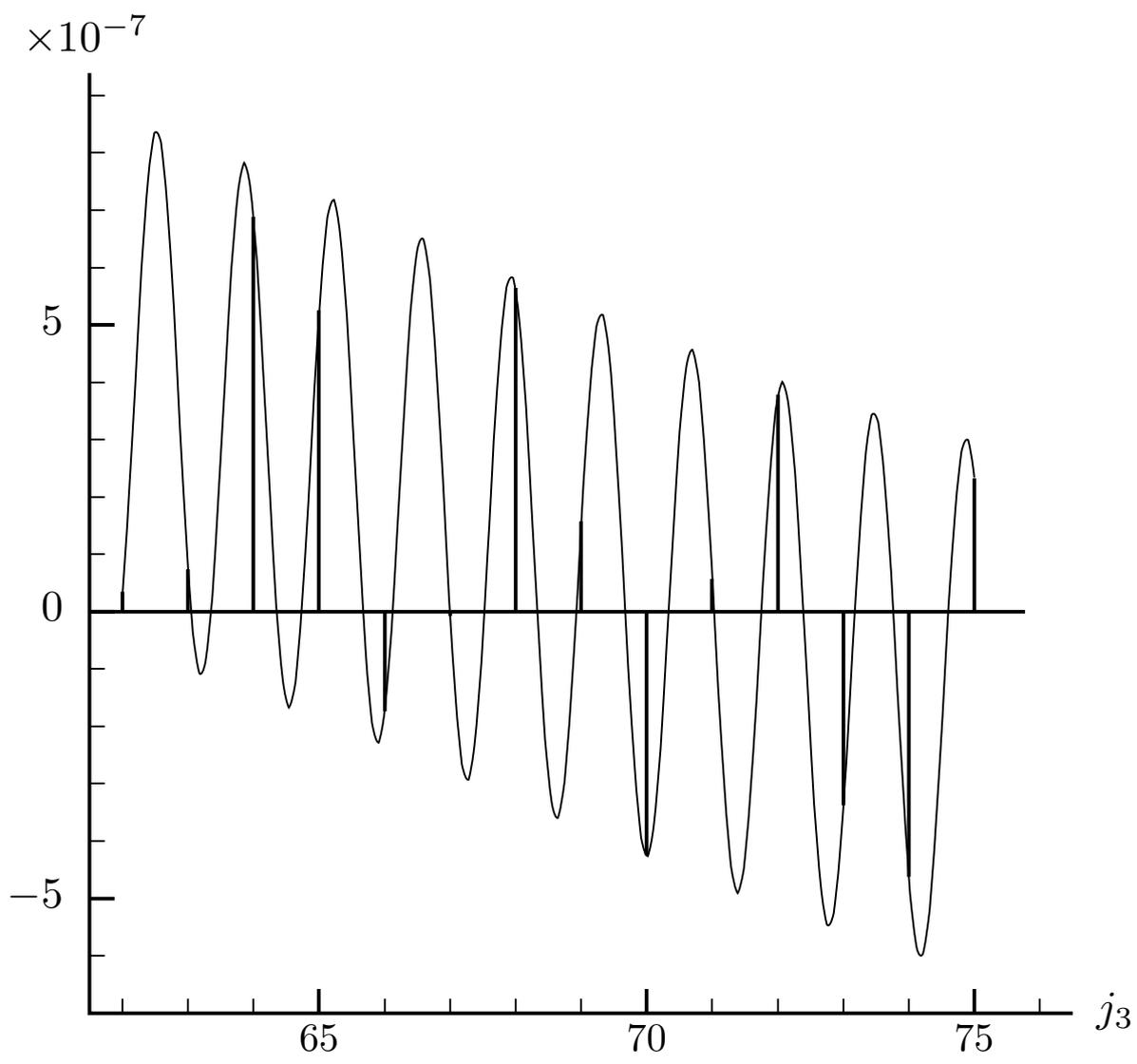
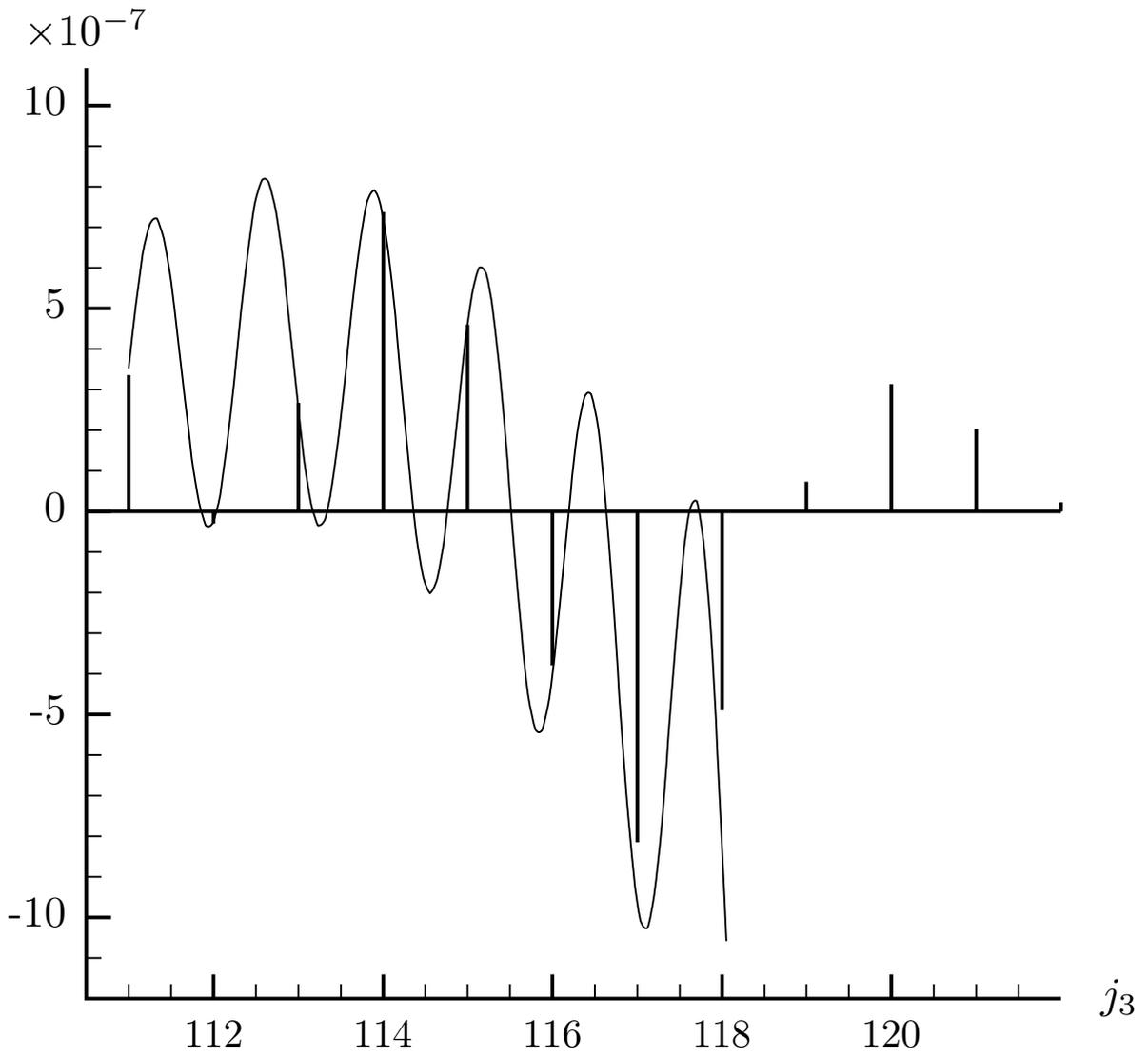
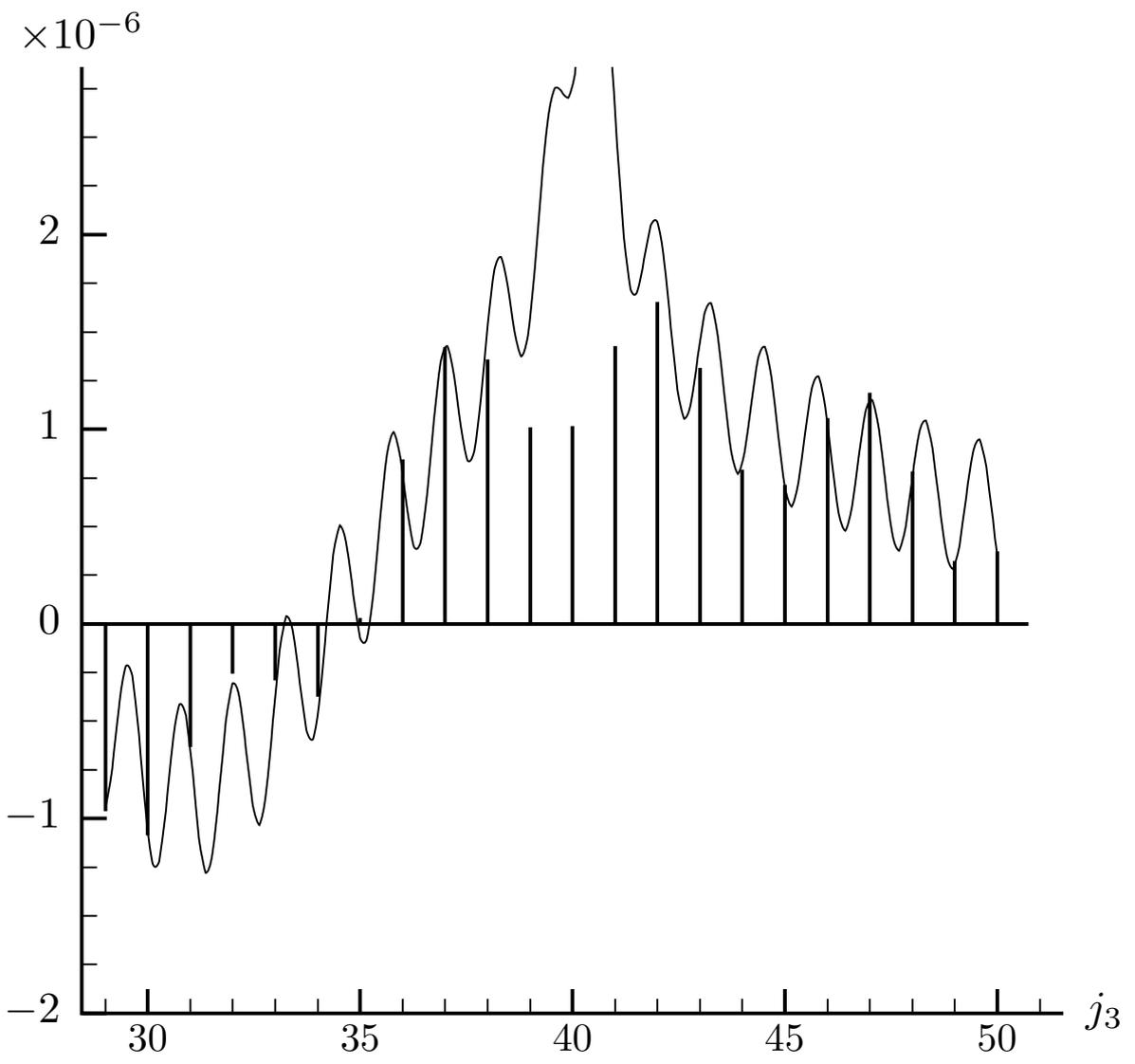
\end{document}